\documentclass[twocolumn,secnumarabic,amssymb,nobibnotes,aps,prl,superscriptaddress,nobalancelastpage,longbibliography,floatfix,noeprints]{revtex4-1}

 \usepackage{graphicx}
\usepackage{bm}
\usepackage{amsmath} \usepackage{braket} \usepackage{epsfig}
\usepackage{tensor}
\usepackage{multirow}
\usepackage[T1]{fontenc}
\usepackage[version=4]{mhchem}
\usepackage{titlesec}
\usepackage{ifthen}
\usepackage{sidecap}
\usepackage{listings}
\usepackage[para,online,flushleft]{threeparttablex}
\usepackage{algorithm}
\usepackage{algpseudocode}
\usepackage{comment}
\usepackage{xr-hyper}
\usepackage{hyperref}
\usepackage[normalem]{ulem}
\usepackage{xcolor}
\usepackage[all]{hypcap}
\hypersetup{breaklinks=true,colorlinks=true,linkcolor=blue,citecolor=blue,filecolor=magenta,urlcolor=cyan}

\definecolor{pastelgray}{rgb}{0.81, 0.81, 0.77}
\definecolor{beaublue}{rgb}{0.9, 0.9, 0.93}

\newcommand{\ham}{\widehat{\mathcal{H}}}
\newcommand{\meff}{\mathcal{M_\textrm{eff}}}
\newcommand{\mten}{\mathcal{M_{\mu\nu}}}
\newcommand{\tsf}{t_{1/2}^\textrm{sf}}
\newcommand{\eqr}[1]{Eq.~\eqref{#1}}

\makeatletter
\def\@bibdataout@aps{%
\immediate\write\@bibdataout{%
@CONTROL{%
apsrev41Control%
\longbibliography@sw{%
    ,author="08",editor="1",pages="1",title="0",year="1"%
    }{%
    ,author="08",editor="1",pages="1",title="",year="1"%
    }%
  }%
}%
\if@filesw \immediate \write \@auxout {\string \citation {apsrev41Control}}\fi
}
\makeatother

\begin{document}

\title{Neural Network Emulation of Spontaneous Fission}

\author{Daniel Lay}
\affiliation{Department of Physics and Astronomy and FRIB Laboratory, Michigan State University, East Lansing, Michigan 48824, USA}

\author{Eric Flynn}
\affiliation{Department of Physics and Astronomy and FRIB Laboratory, Michigan State University, East Lansing, Michigan 48824, USA}

\author{Samuel A. Giuliani}
\affiliation{Departamento de F{\'i}sica Te{\'o}rica and CIAFF, Universidad
Aut{\'o}noma de Madrid, Madrid 28049, Spain}
\affiliation{Department of Physics, Faculty of Engineering and Physical
Sciences, University of Surrey, Guildford, Surrey GU2 7XH, United Kingdom.}

\author{Witold Nazarewicz}
\affiliation{Department of Physics and Astronomy and FRIB Laboratory, Michigan State University, East Lansing, Michigan 48824, USA}

\author{Le{\'o} Neufcourt}
\affiliation{FRIB Laboratory, Michigan State University, East Lansing, Michigan 48824, USA}

\begin{abstract}
\edef\oldrightskip{\the\rightskip}
\begin{description}
\rightskip\oldrightskip\relax
\setlength{\parskip}{0pt}

\item[Background]
Large-scale computations of fission properties are an important ingredient for nuclear reaction network calculations simulating rapid neutron-capture process (the $r$ process) nucleosynthesis. Due to the large number of fissioning nuclei potentially contributing to the $r$ process, a microscopic description of fission based on nuclear density functional theory (DFT) is computationally challenging.

\item[Purpose]
We explore the use of neural networks (NNs) to construct DFT emulators capable of predicting potential energy surfaces and collective inertia tensors across the whole nuclear chart, starting from a minimal set of DFT calculations.

\item[Methods]
We use constrained Hartree-Fock-Bogoliubov (HFB) calculations to predict the potential energy and collective inertia tensor in the axial quadrupole and octupole collective coordinates, for a set of nuclei in the $r$-process region. We then employ NNs to emulate the HFB energy and collective inertia tensor across the considered region of the nuclear chart. Least-action pathways characterizing spontaneous fission half-lives and fragment yields are then obtained by means of the nudged elastic band method.

\item[Results] 
The potential energy predicted by NNs agrees with the DFT value to within a root-mean-square error of 500 keV, and the collective inertia components agree to within an order of magnitude. These results are largely independent of the NN architecture. The exit points on the outer turning line are found to be well emulated. For the spontaneous fission half-lives the NN emulation provides values that are found to agree with the DFT predictions within a factor of $10^3$ across more than 70 orders of magnitude.

\item[Conclusions] 
Neural networks are able to emulate the potential energy and collective inertia well enough to reasonably predict physical observables. Future directions of study, such as the inclusion of additional collective degrees of freedom and active learning, will improve the predictive power of microscopic theory and further enable large-scale fission studies.

\end{description}
\end{abstract}

\date{\today}

\maketitle

\section{Introduction}\label{sec:introduction}
Large scale calculations of fission properties are an essential ingredient for the modelling of the rapid neutron-capture process ($r$ process), responsible for the production of roughly half of the nuclei heavier than iron found in nature~\cite{Burbidge1957,Cameron1957}. Fission determines the range of the heaviest nuclei that can be synthesized during the $r$-process, recycles the material during the neutron irradiation phase, and shapes the final abundances~\cite{Horowitz2018,Kajino2019,Cowan2021}. Given the large amount of energy released in this decay, the presence of fissioning nuclei can leave fingerprints in the electromagnetic counterpart produced in neutron star mergers~\cite{Zhu2018,Wu2018}. However, as most of the fissioning nuclei produced during the $r$ process cannot be measured, theoretical predictions are indispensable to perform accurate nuclear reaction network calculations. 

During the last decades, several efforts have been devoted to the systematic estimation of fission barriers \cite{Aboussir1992,Myers1999,Goriely2009,Moller2015,Agbemava2017,Giuliani2018}, spontaneous fission half-lives \cite{Staszczak2013,Baran2015,Giuliani2018}, and fragment distributions~\cite{Kelic2009,SCHMIDT2016107,Mumpower2020,Lemaitre2021,Sadhukhan2022} of $r$-process nuclei. However, due to the inherent complexities characterizing the theoretical description of the fission process~\cite{Bender2020}, most of the available calculations resort to phenomenological approaches based on simplified assumptions. This limitation can be overcome by employing the nuclear DFT~\cite{Bender2003,Duguet2014,Schunck2019a}, which is the quantum many-body method based on effective nucleon-nucleon interactions applicable across the whole nuclear landscape. But given its computational costs, using DFT for fission is a daunting task for large-scale studies of $r$-process nuclei~\cite{schunck2016,Bender2020,Schunck2022}. As such, the usage of DFT emulators can be an invaluable tool to extend the current reach of microscopic fission calculations.

Machine learning has been used with great success in many areas of nuclear physics (see~\cite{MLinNuclearPhysics} for a recent review on this topic). In particular, machine learning has been used in many DFT studies to emulate potential energy surfaces (PESs), in both quantum chemistry~\cite{Manzhos2021,Nagai2020,Bin2020,Dral2020} and in nuclear physics~\cite{Akkoyun2013,Verriere2022}. However, these have generally focused on emulating individual potential energy surfaces, rather than many nuclei across a portion of the nuclear chart (or many related chemical systems in the quantum chemistry case). In an important study, Ref.~\cite{Lasseri2020} succeeded in emulating PESs and other  quantities using  committees of multilayer neural networks.

In this study, we use fully connected, feedforward NNs to emulate the PES and collective inertia tensor, parameterized by the axial quadrupole and octupole moments $Q_{20}$ and $Q_{30}$, between nuclei in the $r$-process region of the nuclear chart. The paper is organized as follows: Section~\ref{sec:dft} reviews the theoretical approach to spontaneous fission used in this work. Section~\ref{sec:nn} describes the characteristics of the employed NNs. Section~\ref{sec:nnquality} demonstrates the performance of the NNs on the HFB energy and collective inertia tensor, and Sec.~\ref{sec:observables} compares the exit points and spontaneous fission half-lives obtained using the DFT inputs and the emulated NN inputs. Finally, conclusions are summarized in Sec.~\ref{sec:conclusion}.

\section{Spontaneous fission within the nuclear density functional theory}
\label{sec:dft}
Spontaneous fission (SF) is a dynamical process where the nucleus evolves from the ground-state into a split configuration. In the adiabatic approximation, SF is modeled using a finite set of collective variables $\{{q_i}\}$ usually describing the nuclear shape. The SF half-life can be computed within this approach as $t_{1/2} = {\ln 2}/{nP_\textrm{fis}}$, where $n$ is the number of assaults on the fission barrier, and $P_\textrm{fis}$ the fission probability given by the probability of the nucleus to tunnel through the fission barrier, which can be estimated using the semiclassical Wentzel–Kramers–Brillouin (WKB) approach~\cite{Brack1972}:
\begin{equation}\label{eq:pfis}
    P_\textrm{fis} = \frac{1}{1+\exp \left( 2S(L) \right)},
\end{equation}
where $S(L)$ is the collective action computed along the stationary trajectory $L[s]$ that minimizes $S$ in the multidimensional space defined by the collective coordinates:
\begin{equation}\label{eq:actionint}
    S(L[s]) = \frac{1}{\hbar} \int_{s_\textrm{in}}^{s_\textrm{out}} \sqrt{2\meff(s)(V(s)-E_0)} \,\, ds\,,
\end{equation}
with $V$ and $\meff$ being the potential energy and inertia tensor, respectively, computed along the fission path $L[s]$. The integration limits $s_\textrm{in}$ and $s_\textrm{out}$ correspond to the classical inner and outer turning points, respectively, defined by the condition $V=E_0$, where $E_0$ is the collective ground-state zero-point energy stemming from quantum fluctuations in the collective coordinates. While the latter can be estimated from, e.g., the curvature of $V$ around the ground state (g.s.) configuration, in many SF studies $E_0$ is taken as a fixed positive constant ranging between 0.5 and 2.0~MeV above the ground-state energy~\cite{Baran2015,Giuliani2018}. For simplicity, we follow the latter approach and fix $E_0= E_{\textrm{g.s.}}$. And, throughout this work, we will refer to the collective coordinates at $s_\textrm{out}$ (in this work, $(Q_{20},Q_{30})$) as the exit point \cite{flynn2022nudged}.

From~\eqr{eq:actionint} it can be deduced that the main ingredients required for the estimation of the SF half-lives are the effective potential energy $V$ and collective inertia $\meff$. In this work, we compute these quantities by employing the self-consistent mean-field method~\cite{Bender2003,Schunck2019a} summarized in the following. 

Nuclear configurations are obtained by means of the HFB method, where the many-body wave function $|\Psi\rangle$, described as a generalized quasiparticle product state, is given by the minimization of the mean value of the Routhian:
\begin{equation}\label{eq:rout}
    \ham' =
    \ham_\textup{HFB} - \sum_{\tau=n,p} \lambda_\tau \widehat{N}_\tau - \sum_{\mu=1,2,3} \lambda_{\mu} \widehat{Q}_{\mu0}
    \,.
\end{equation}
In~\eqr{eq:rout}, $\ham_\textup{HFB}$  is the HFB Hamiltonian, and $\lambda_p$ and $\lambda_n$ are the Lagrange multipliers fixing the average number of protons and neutrons, respectively. The shape of the nucleus is enforced by constraining the moment operator $\widehat{Q}_{\mu\nu}$ with multipolarity $\mu$ and magnetic quantum number $\nu$. In this work, we explore the evolution of the total energy and collective inertia tensor as a function of the elongation of the nucleus and its mass asymmetry, which are described by the axial quadrupole $Q_{20}$ and octupole $Q_{30}$ moment operators, respectively:
\begin{subequations}
    \begin{align}
        \widehat{Q}_{20} &= \hat{z}^2 - \frac{1}{2} (\hat{x}^2 + \hat{y}^2)\,; \\
        \widehat{Q}_{30} &= \hat{z}^3 - \frac{3}{2} (\hat{y}^2 + \hat{x}^2)\hat{z}\,.
    \end{align}
\end{subequations}
In order to reduce the computational cost, axial symmetry is enforced in all the calculations ($\langle \widehat{Q}_{\mu\nu} \rangle = 0$ for all $\nu \neq 0$), and the additional constraint $\langle \widehat{Q}_{10} \rangle = 0$ is imposed to remove the spurious center-of-mass. Finally, the nuclear HFB Hamiltonian $\ham_\textup{HFB}$ is given by the finite-range density-dependent nucleon-nucleon Gogny interaction. We employ the D1S parametrization~\cite{berger1984microscopic}, which has been widely used in nuclear structure studies across the whole nuclear chart, including the description of fission properties of heavy and superheavy nuclei~\cite{Robledo2018}. The effective potential is then given by $V = E - E_\textrm{rot}$, where $E$ is the energy obtained from the HFB equations for the Routhian \eqref{eq:rout}, and $E_\textrm{rot}$ is the energy correction related to the restoration of rotational symmetry, computed using the approach of Ref.~\cite{Egido2004}. Calculations are carried out by employing the HFB solver HFBaxial, which solves the HFB equations by means of a gradient method with an approximate second-order derivative~\cite{Robledo2011d}. The quasiparticle wave functions are expanded in an axially-symmetric deformed harmonic oscillator single-particle basis, containing states with $J_z$ quantum number up to $35/2$ and up to 26 quanta in the $z$-direction. The basis quantum numbers are restricted by the condition $ 2 n_\perp + |m| + {n_z}/{q} \leq N_z^\mathrm{max}$, where $q=1.5$ and $N_z^\mathrm{max}=17$. This choice of the basis parameters allows for a proper description of the elongated prolate shapes characteristic of the fission process~\cite{Warda2002}.

The collective inertia tensor $\mten$ is computed within the Adiabatic-Time-Dependent HFB (ATDHFB) approximation using the non-perturbative scheme~\cite{Yuldashbaeva1999,Baran2011,Giuliani2018b}:
\begin{equation}
    \mten = \frac{\hbar^2}{2 \dot{q}_\mu \dot{q}_\nu} \sum_{\alpha\beta}
    \frac{F^{\mu*}_{\alpha\beta} F^{\nu}_{\alpha\beta} + F^{\mu}_{\alpha\beta} F^{\nu*}_{\alpha\beta}}{E_\alpha + E_\beta} \,,
\end{equation}
where $q_i$ are the collective coordinates and
\begin{equation}\label{eq:matf}
    \frac{F^\mu}{\dot{q}_\mu} = 
    A^\dagger \frac{\partial \rho}{\partial q_\mu} B^*
    +
    A^\dagger \frac{\partial \kappa}{\partial q_\mu} A^*
    -
    B^\dagger \frac{\partial \rho^*}{\partial q_\mu} A^*
    -
    B^\dagger \frac{\partial \kappa^*}{\partial q_\mu} B^*
\end{equation}
is given in terms of the matrices of the Bogoliubov transformation $A$ and $B$, and the corresponding particle $\rho$ and pairing $\kappa$ densities. Then, the effective inertia tensor is given as
\begin{equation}
    \meff = \sum_{\mu \nu} \mten \frac{dq_\mu}{ds} \frac{dq_\nu}{ds}\,.
\end{equation}
It is important to remark that the $\mten$ components can suffer from rapid oscillations in the presence of single-particle level crossings near the Fermi surface. Such abrupt changes of occupied single-particle configurations produce variations in the derivatives of the densities in~\eqr{eq:matf}, resulting in pronounced peaks of $\meff$ along the fission path~\cite{Baran2011,Sadhukhan2013,Giuliani2018b}. 

The least action paths are computed using the nudged elastic band method (NEB)~\cite{flynn2022nudged}. Due to the large number of paths that must be explored, NEB parameters cannot be tuned by hand. Instead, multiple NEB runs are started, with initial paths ending at various points along the outer turning line. The NEB algorithm depends on two parameters, $k$ and $\kappa$, which adjust spring and harmonic restoring forces, respectively. Not varying $k$ and $\kappa$ will, on occasion, miss some LAPs, akin to skipping over a narrow minimum in an optimization routine. Different runs are started for $k$ and $\kappa$ in the range $0.05-10$, for each initial path. These runs converge to a number of different stationary paths. Typically, there is some component of the path that travels along the outer turning line. To select the final tunneling path, the paths are interpolated using 500 points, and truncated when near the outer turning line and within an energy tolerance of $0.5$ MeV. The unique paths are chosen based on the clustering of the exit point using the mean shift algorithm as implemented in scikit-learn~\cite{scikit-learn}, and the path corresponding to a given exit point with the least action is chosen as the LAP.

\section{Neural networks}
\label{sec:nn}

In this work, we use feedforward NNs as our emulators. We train separate NNs on the potential energy $V$ and the components of $\mathcal{M}$. Each NN takes as input $(A,Z,Q_{20},Q_{30})$, specifying the nucleus and deformation, then outputs the value (either $V$ or one of $\mathcal{M}_{\mu\nu}$) at that point. As discussed in Sec.~\,\ref{subsec:pes-nn}, to further improve NN performance, we rescale the NN inputs to lie between zero and one (to avoid biasing the NNs, the same linear rescaling factors are used for all data points).

We train NNs with a number of hidden layers varying between 2 and 7, with 200 hidden nodes in the first layer, and a decreasing number of nodes in each subsequent layer. We use the RELU activation function, and train to minimize the root-mean-square error in the desired quantity. For each variant on the NN depth, we train multiple NNs, forming a committee of NNs. We then combine the predictions from each NN in the committee in a weighted average, to further reduce the error on the prediction.

To train the NNs, we have computed PESs and the collective inertia for 194 nuclei, each on a regular grid of 4~b for $0\leq Q_{20}\leq 248$ b, and 6 b$^{3/2}$ for $0\leq Q_{30}\leq 60$ b$^{3/2}$. These nuclei are then labeled as either training, combining, or validation. The combining and validation nuclei were sampled from a uniform random distribution, such that no region of the nuclear chart is overrepresented in either dataset; the remaining nuclei form the training set. These different datasets are indicated in Fig.~\ref{fig:rmse_eneg}. For each nucleus, the entire grid is used in the training/combining/validation. The nuclei in the training set are used to train individual NNs, the nuclei in the combining dataset are used to combine predictions from the committee members in a weighted average, and the nuclei in the validation set are used for validation of the NN predictions. The weights for each committee member are chosen to minimize the root-mean-square error on the nuclei in the combining dataset. As can be seen, most of the nuclei (about 70\%) are used for training, with the remaining 30\% split equally between the combining and validation datasets. In general, the NN performance is not sensitive to the distribution of training data, provided the NN does not attempt to extrapolate across the nuclear chart. No detailed optimization of the choice of training nuclei was carried out.

As mentioned in Sec.~\ref{sec:dft}, it is known that the collective inertia tensor can develop discontinuities and rapid variations due to level crossings. This makes emulation of the tensor challenging since the tensor components can span many orders of magnitude as a function of deformation. If the NN is trained on the inertia tensor components by themselves, the network predictions are poor. However, while these problems are features of the approximations used to calculate the inertia, the NN can still learn certain features of the inertia tensor by carrying out the eigenvalue decomposition of the inertia tensor,
\begin{align}
    \mathcal{M} = U \Sigma U^{T}
\end{align}
where $U$ is the $2 \times 2$ matrix of eigenvectors and $\Sigma$ is the diagonal matrix of eigenvalues. Since $U$ is an orthogonal matrix, we can represent $U$ as an element of the set $\textrm{SO}(2)$ parameterized by Euler angle $\theta$. In this representation, $\mathcal{M}$ is completely parameterized by its eigenvalues and the Euler angle $\theta$. So, the NN is trained on $\theta$ and the log of the eigenvalues at each point ($Q_{20}$, $Q_{30}$). Training on this representation of the tensor is similar to normalizing the network inputs, as both put NN inputs/outputs on a similar scale. Additionally, this forces the tensor predictions to be positive semi-definite. We also transform $\theta$ to the range $(-\pi/2,\pi/2)$, so that the angles are mostly clustered near zero (on the interval $(0,\pi)$, there are two clusters: one at 0 and one at $\pi$, which the NN has difficulties learning). 

Once the NNs are trained, PESs and inertias are computed for the same grid of deformations as the original DFT calculations. While the NNs can be evaluated at arbitrary $(Q_{20},Q_{30})$, it is less computationally expensive to use a standard cubic spline interpolator on the grid predicted by the NN. Moreover, the LAPs computed using the NN evaluations and the spline interpolator agree well with each other. Due to the relatively large number of LAP calculations, we report the LAPs computed using the spline-interpolated NN predictions, rather than using the NN predictions themselves.

\section{Neural Network Quality}
\label{sec:nnquality}
Here, we examine the quality of the NNs, on both the PES and the collective inertia. In general, we observe that the NN is able to reproduce both the PES and the collective inertia for most of the nuclei under consideration. Moreover, the quality of the NN is relatively stable across the different architectures considered.

Throughout this section, we will refer to the PES and collective inertia computed using DFT as the reference data, and the PES and inertia computed using the NN as the NN reconstruction.

\subsection{Potential energy surfaces}
\label{subsec:pes-nn}
For a single nucleus, we define the root-mean-square error (RMSE) $\Delta V(A,Z)$ in energy over the collective domain considered  as  
\begin{align}
    \Delta V(A,Z)^2=\frac{1}{n}\sum_{Q_{20},Q_{30}}[&V^{\textrm{DFT}}(Q_{20},Q_{30},A,Z)\nonumber\\
    &\,\,-V^{\textrm{NN}}(Q_{20},Q_{30},A,Z)]^2,
\end{align}
where $n=693$ is the number of gridpoints evaluated in the PES. A similar quantity can be defined for the components of $\meff$, although there $n$ varies slightly from nucleus to nucleus. 

\begin{figure}
\label{fig:rmse_eneg}
    \includegraphics[width=\linewidth]{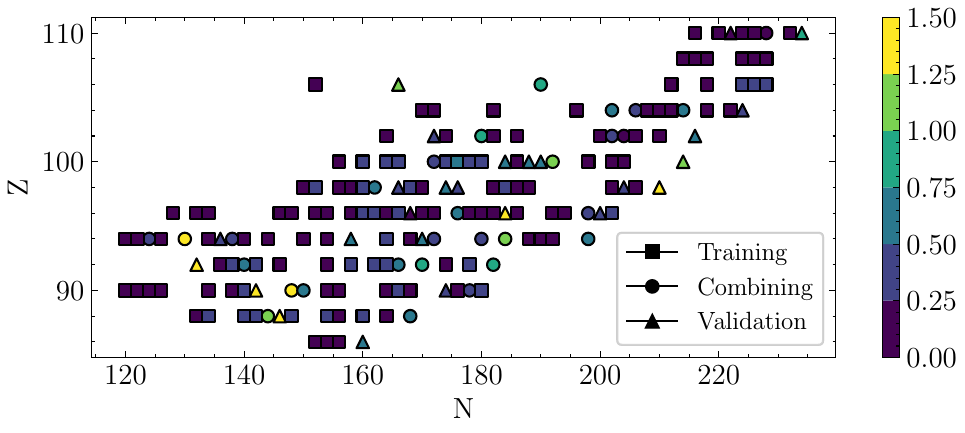}
    \caption{$\Delta V(A,Z)$ (in MeV) for the deepest NN. The different shapes indicate which dataset each nucleus belongs to.}
\end{figure}

Figure~\ref{fig:rmse_eneg} shows $\Delta V(A,Z)$ across the region of the nuclear chart considered, for the deepest NN (7 hidden layers, with 200-175-150-125-100-75-50 hidden units), with rescaled inputs. As can be seen, for most nuclei, $\Delta V(A,Z)\lesssim 0.5$~MeV. Exceptions occur, with most remaining below 1.5~MeV.
For some nuclei, such as $^{308}$Cf, $^{314}$Fm, and $^{318}$No, relatively poor performance may be expected: these nuclei are on the outer edge of the region of the nuclear chart considered, and hence the NN is extrapolating from the training region to reach them. For other nuclei, such as $^{232}$Th and $^{280}$Cm, poor performance is unexpected: these nuclei are surrounded by training nuclei, and so should be emulated fairly well. As such, it seems unlikely that poor performance is due solely to the location of the nucleus on the nuclear chart relative to the training data.

\begin{figure}[htb]
        \includegraphics[width=\linewidth]{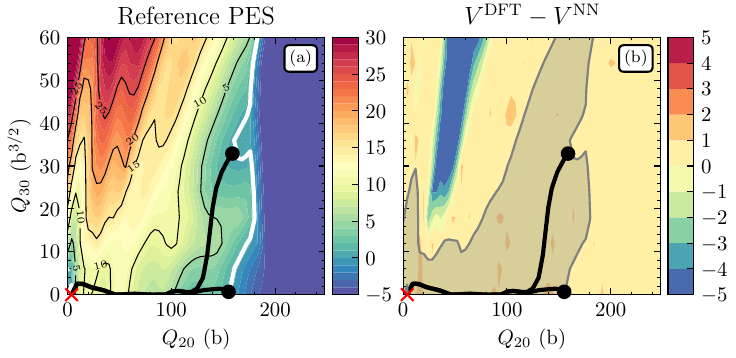}
        \caption{(a): the reference PES for $^{280}$Cm in MeV. (b): the difference between the reference and reconstructed PESs, again for $^{280}$Cm, in MeV. The reference ground state and the reference LAPs are marked with a $\color{red}\bm{\times}$ symbol and black lines, respectively, in both panels. The energy range $0.5\leq V^{\textrm{DFT}}\leq 12$ MeV is shaded in gray in panel (b).}
        \label{fig:280Cm}
\end{figure}

To understand the reduced performance, we examine the nuclei in question. Figure~\ref{fig:280Cm}(a) shows the reference PES for ${}^{280}$Cm, and Fig.~\ref{fig:280Cm}(b) shows the difference between the reference PES and its NN reconstruction. This nucleus is chosen because it has $\Delta V=2.15$~MeV, which is the largest of all nuclei in the validation set. The difference is less than 1 MeV across most of the PES, including in the region relevant for fission. The energy difference is large elsewhere, with a difference of more than 5 MeV, which is why $\Delta V$ is rather large for $^{280}$Cm. We conclude that even for nuclei with larger RMSE, NNs could provide a very reasonable description of the fission path. This aspect will be examined further in Sec.~\ref{sec:observables}.

\begin{figure}[htb]
    \includegraphics[width=\linewidth]{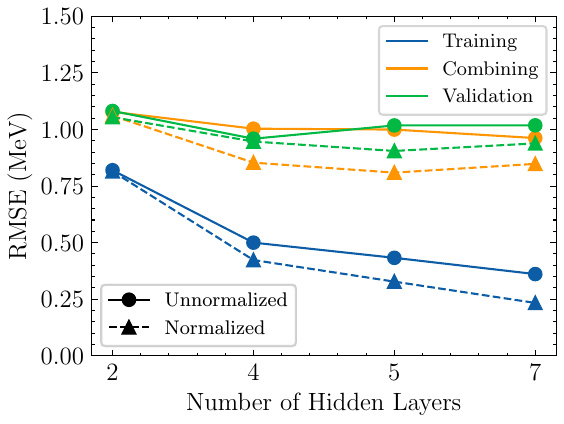}
    \caption{The RMSE for a variety of different NN sizes. The dashed line shows the same depth NN, but with input variables normalized to the range $[0,1]$.}
    \label{fig:size-variation-rmse}
\end{figure}

To assess the sensitivity of our results with respect to the NN architecture, we repeated our calculations employing different NN sizes and rescaling the inputs. Figure~\ref{fig:size-variation-rmse} shows the RMSE (now averaged across all nuclei in a given dataset) across the different datasets for a variety of NN depths. As is generally expected, the training dataset has a monotonically decreasing RMSE as the NN depth increases; this is simply due to the increasing number of tunable parameters in the NN. On the other hand, the RMSE for the combining and validation sets is fairly stable with respect to the number of hidden layers of the NN. 

A general improvement is observed when normalizing the inputs $(A,Z,Q_{20},Q_{30})$ to be between 0 and 1. This is due to two factors. First, the weights are not scale-invariant. An input much larger than 1 is equivalent to a large initial weight, with a normalized input. Because the final NN weights are expected to be small (hence initializations following e.g. the Xavier initialization~\cite{Xavier-init}, as in this work), the initial weights are far from the final values, and convergence slows. Indeed, the final NN weights are small, and the distribution of the weights is similar when comparing NNs with normalized and non-normalized inputs. Second, the optimization method itself is not scale-invariant: non-normalized inputs correspond to an ill-conditioned Hessian matrix, in which case gradient descent (and related methods) converge slowly~\cite{Curry1944TheMO,RIGLER1991225}.

We conclude that the NN performance in predicting the PES is relatively stable with respect to the NN architecture; Sec.~\ref{sec:observables} will demonstrate that performance on this level is adequate for predicting SF observables.

\subsection{Collective Inertia}
\begin{figure}
    \includegraphics[width=\linewidth]{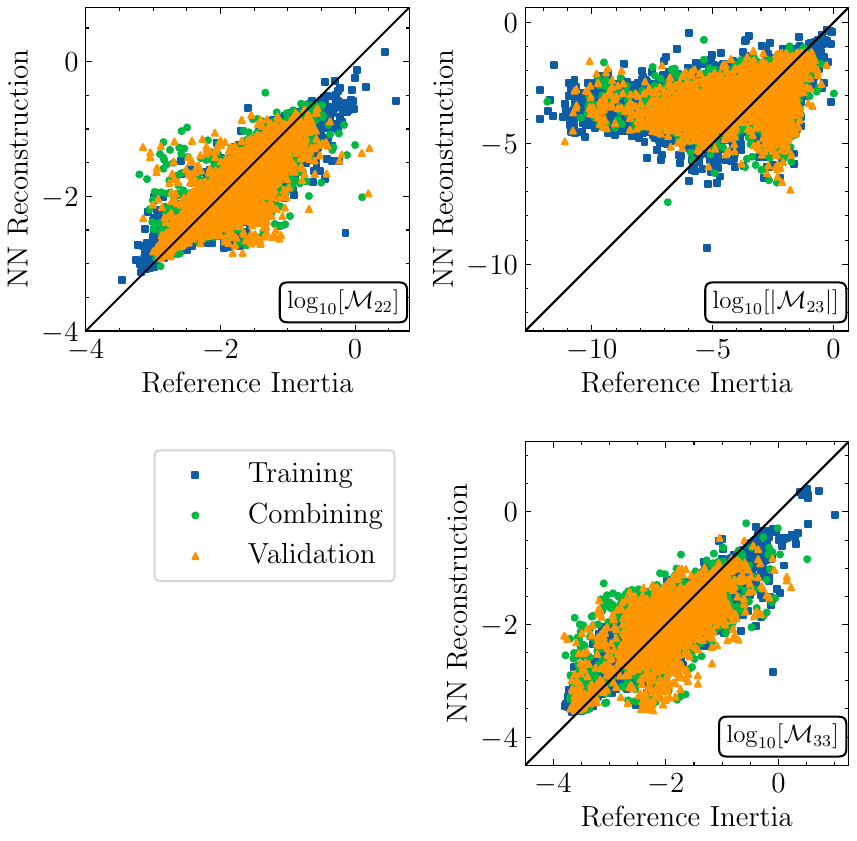}
    \caption{The reference components of $\mathcal{M}$, plotted against the NN reconstructions, for all nuclei considered. The black line is the diagonal, $\mathcal{M}_{\mu\nu}^{\textrm{DFT}}=\mathcal{M}_{\mu\nu}^{\textrm{NN}}$. The blue squares/green circles/orange triangles correspond to the training/combining/validation datasets. Note that, for use on a log scale, we plot the absolute value of $\mathcal{M}_{23}$ (the other components are nonnegative). $\mathcal{M}_{22}$ is in 
    MeV$^{-1}$b$^{-2}$, $\mathcal{M}_{23}$ is in 
    MeV$^{-1}$b$^{-5/2}$, and $\mathcal{M}_{33}$ is in
    MeV$^{-1}$b$^{-3}$.} 
    \label{fig:inertia-diag}
\end{figure}

Since the components of $\mathcal{M}$ vary across multiple orders of magnitude and the network is trained on the log of the eigenvalue decomposition, a loss function such as the root-mean-squared error is not an adequate measure of the performance of the NN. Instead, Fig.~\ref{fig:inertia-diag} shows the reference inertia components, plotted against the NN reconstructions, for all nuclei considered. The NN used is the 7-hidden-layer NN with rescaled inputs, with the number of hidden units as described in Sec.~\ref{subsec:pes-nn}. The diagonal components $\mathcal{M}_{22}$ and $\mathcal{M}_{33}$ are predicted fairly well, as the distributions align roughly along the diagonal. It is worth noting that the distributions are slightly misaligned, in all data sets considered, indicating that the NN tends to underpredict relatively large values, and overpredict relatively small values. This, in turn, shows that the NN is slightly biased towards the mean value of the inertia.

However, the off-diagonal component $|\mathcal{M}_{23}|$ is not aligned along the diagonal, except for large values. This is because this component actually varies across almost 10 orders of magnitude (compared to the 4 orders of magnitude for $\mathcal{M}_{22}$ and $\mathcal{M}_{33}$), and so the NN is biased towards predicting the larger values more accurately, resulting in a general overprediction of $\mathcal{M}_{23}$. In terms of the angle $\theta$ that is actually determined by the NN, it is difficult to predict both small and large angles, and because $\theta$ is allowed to be negative, a logarithm transform is not possible. Nevertheless, one obtains a reasonable-looking distribution above $|\mathcal{M}_{23}|\gtrsim 10^{-4}$\,MeV$^{-1}$\,b$^{-5/2}$, indicating that some learning has indeed taken place. And, the poorly-learned values below $10^{-4}$\,\,MeV$^{-1}$\,b$^{-5/2}$ are truncated at values $10^{-6}-10^{-2}$\,\,MeV$^{-1}$\,b$^{-5/2}$. 

When changing the depth of the NN, performance is similar. For shallow networks, predictions on the training dataset show a larger bias: the distribution of points on the inertia plot is less aligned with the diagonal for the $\mathcal{M}_{22}$ and $\mathcal{M}_{33}$ components. In other words, the larger reference values are underestimated, and the smaller reference values are overestimated. The validation dataset is aligned similar to the deepest network, shown in Fig.~\ref{fig:inertia-diag}. As the depth of the network is increased, the training data points are aligned closer with the diagonal. This is indicative of the NN tending to overfit on the training data as the number of variational parameters increases. The distribution of $\mathcal{M}_{23}$ values remains approximately the same when increasing NN depth, with a slight improvement on the truncated $\mathcal{M}_{23}$ values. In general, the NN performance on the validation dataset is mostly stable when varying the NN depth. The overarching question is whether this performance is sufficient for predicting observable quantities of interest. As with the PES, this question can be directly answered by looking at NN predictions of physical observables.

\section{Impact on Observable Quantities}
\label{sec:observables}
While encouraging, the results discussed in Sec.~\ref{sec:nnquality} do not give a perfectly clear estimation of the performance of the NNs. For instance, the NN reconstruction of the PES for $^{280}$Cm may be adequate for reproducing fission observables - especially SF fragment yields and half-lives, despite the poor RMSE, since the largest deviations occur at deformations that will not be explored by LAPs. Similarly, the NN commonly fails to reproduce the off-diagonal component of the collective inertia, $\mathcal{M}_{23}$, but primarily for small values of $\mathcal{M}_{23}$. 

Here, we examine the performance of the NN on the lifetime-weighted exit point, as a proxy for the fragment yield~\cite{Sadhukhan2020,Sadhukhan2022}, and the half-life of the nucleus. For both quantities, we compare three sets of data: the quantity computed using (i) the reconstructed PES and the identity inertia; (ii) using the reference PES and the reconstructed inertia; and (iii) the reconstructed PES and inertia. In this way, we can isolate the impact of the PES and inertia emulations separately, and combine them to assess the overall error of the emulator. In this section, we use the 7-hidden-layer NN with rescaled inputs, with a number of hidden units described in Sec.~\ref{subsec:pes-nn}. Based on the relative insensitivity to the depth of the NN shown in Sec.~\ref{sec:nnquality}, the overall performance is expected to be similar for different NN depths.

Similar to Sec.~\ref{sec:nnquality}, exit points and SF half-lives computed using only DFT inputs will be referred to as reference quantities; those with any NN input will be referred to as reconstructed quantities.

\subsection{Exit Points}\label{subsec:exit-points}
As demonstrated in~\cite{Sadhukhan2020,Sadhukhan2022}, the location of the exit points is sufficient for roughly estimating the fission fragment yields. For this reason, we can consider exit points as reasonable proxies for the fragment yields. When multiple fission channels exist, the combined fragment yields are attained by adding the yields of each channel, weighted by the probability of populating a particular channel.  Thus, agreement of the lifetime-weighted exit point indicates strong agreement in the fission fragment yields (and, by necessity, indicates that the dominant fission mode is also in agreement between the reference data and the NN reconstruction).

\begin{figure}[htb]
    \centering
    \includegraphics[width=\linewidth]{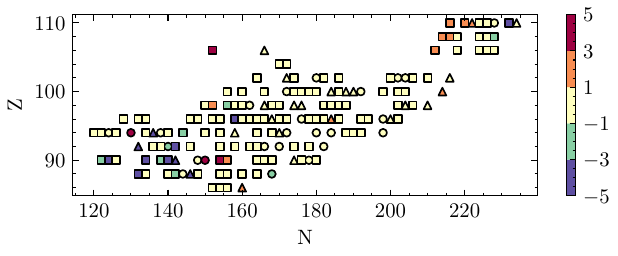}
    \caption{The $Q_{30}$ component of the reconstructed lifetime-weighted exit point, minus the reference $Q_{30}$ component, in b$^{3/2}$. These results were computed using configuration (iii), i.e. the NN was used to reconstruct both the PES and the collective inertia.}\label{fig:otp-diff}
\end{figure}

Figure~\ref{fig:otp-diff} shows the difference in the octupole moment of the lifetime-weighted exit point, for configuration (iii) mentioned above. The octupole moment is chosen because it is critical for explaining multimodality in SF. The $Q_{30}$ error is similar for the other configurations, and the quadrupole moment is typically within $\pm1$\,b for all configurations.

The agreement is good between the reference exit point and the NN reconstruction: at $\pm1$~b$^{3/2}$, we expect the fragment yields to agree well (within the hybrid method of Refs.~\cite{Sadhukhan2020,Sadhukhan2022}). This agreement is mainly due to the accurate PES reconstruction, as previous studies have shown that the exit point location is fairly robust with respect to variations in the collective inertia~\cite{Sadhukhan2013,Sadhukhan2014,Matheson2019,Sadhukhan2020}. This agreement holds even for nuclei whose PES reconstruction has a large error, such as $^{280}$Cm, indicating that the qualitative features shown in Fig.~\ref{fig:280Cm}(a) are reconstructed well enough to describe multimodality in SF.

Notice, however, that the exit point locations are not reproduced perfectly for some nuclei, especially in the thorium ($Z=90$) chain, where the difference can be as much as 5\,b$^{3/2}$. This is not due to the PES reconstruction: Fig.~\ref{fig:rmse_eneg} shows that the thorium isotopes have RMSE $\Delta V(Z=90)\lesssim100$~keV, and the exit point reconstruction when considering configuration (i) is within $1$\,b$^{3/2}$ of the reference value. Additionally, a side-by-side comparison of the collective inertia components does not show a systematic deviation between the reference inertia and the NN reconstruction.

Nevertheless, the error is due to the inaccurate collective inertia reconstruction. However, it is not a systematic error. Rather, random error is present for every deformation considered, and it is the accumulation of this random error that causes the discrepancy. While the location of any individual exit point is not sensitive to the collective inertia, the probability of tunneling to a particular point depends on the probability given in Eq.~(\ref{eq:pfis}). Because the probability is exponentially dependent on the action (and therefore exponentially dependent on the collective inertia reconstruction), comparatively small errors can add up and actually switch the dominant exit point, from asymmetric to symmetric and vice versa. This is especially important for nuclei with a wide fission barrier, as the cumulative error along the path is large.

In general, we observe that both the PES and the collective inertia are emulated well enough to predict exit points that agree with the reference data. And, for most nuclei, the dominant mode is also in agreement. Together, this means that the SF fragment yields are in agreement between the reference data and the NN reconstruction for most nuclei under consideration.

\subsection{Spontaneous fission half-lives}
In this section we examine the performance of the NN when predicting the SF half-life, $\tsf$. For the sake of simplicity, we do not include triaxiality and pairing correlations as collective degrees of freedom, despite their large impact on the predicted $\tsf$~\cite{Giuliani2013,Rodriguez-Guzman2014a,Sadhukhan2014,Giuliani2014,Zhao2016}.

\begin{figure}[htb]
    \includegraphics[width=\linewidth]{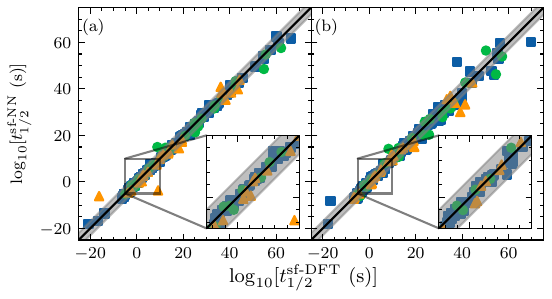}
    \caption{The half-life predicted using the DFT reference data, $t_{1/2}^\textrm{sf-DFT}$, plotted against the half-life computed using the NN reconstruction, $t_{1/2}^\textrm{sf-NN}$. Panel~(a) shows configuration (i), in which only the PES is emulated; panel~(b) shows configuration (iii), in which both the PES and the collective inertia are emulated. The black line marks the diagonal: $t_{1/2}^\textrm{sf-DFT}=t_{1/2}^\textrm{sf-NN}$. Gray bars are drawn at $t_{1/2}^\textrm{sf-DFT}\times 10^{\pm3}$, i.e. 3 orders of magnitude above and below the diagonal. Insets  show the range $10^{-5}-10^{10}$\,s, to highlight the relevant $r$-process range.}
    \label{fig:lifetime-diag}
\end{figure}

Figure~\ref{fig:lifetime-diag} shows $\tsf$ computed using the reference data vs. $\tsf$ computed using the NN reconstruction, for configurations (i) and (iii) mentioned above (results for 
configuration (ii) are similar to those of (iii)). As can be seen, the $\tsf$ predictions agree well, typically within 3 orders of magnitude across the approximately 80 orders of magnitude under consideration.

Figure~\ref{fig:lifetime-diag}(a) demonstrates that the PES reconstruction is sufficient to predict $\tsf$ values that agree well with the reference values. As with the SF fragment yields, this is true even for nuclei with a large $\Delta V$, e.g. $^{280}$Cm, once again demonstrating that the PES emulation quality is indeed sufficient to reproduce SF observables.

Figure~\ref{fig:lifetime-diag}(b) includes the collective inertia emulation. As can be seen, the reproduced $\tsf$ values agree less well with the reference values, although the disagreement is still within 3 orders of magnitude for most nuclei. This is not unexpected: Sec.~\ref{subsec:exit-points} shows that the collective inertia emulation, while sufficient for most nuclei, is not accurate enough for all nuclei.

\begin{figure}[htb]
    \includegraphics[width=\linewidth]{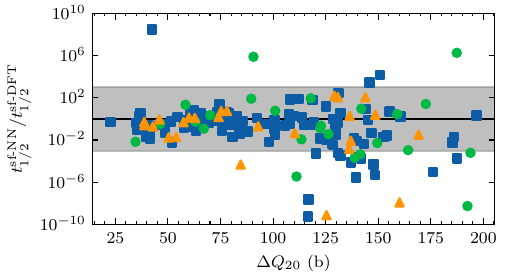}
    \caption{The ratio of the reconstructed half-life, $t_{1/2}^{\textrm{sf-NN}}$, to the reference half-life, $t_{1/2}^{\textrm{sf-DFT}}$, plotted against the difference in $Q_{20}$ between the ground state and the exit point, for configuration (iii). Gray bars mark a three orders of magnitude region.}
    \label{fig:lifetime-error}
\end{figure}

Similar to Sec.~\ref{subsec:exit-points}, the reason for the disagreement in $\tsf$ is the accumulation of random errors when the fission pathway goes across the fission barrier. Now, rather than changing the dominant fission mode, $\tsf$ is simply changed from the reference value in a more-or-less random manner. The effect is most prominent for long-lived nuclei, where errors in the collective inertia add up to a fairly large value as the pathway traverses a wider fission barrier. This is demonstrated in Fig.~\ref{fig:lifetime-error}, where the ratio of the half-lives is plotted against the barrier width, defined here as the difference between $Q_{20}$ at the exit point and $Q_{20}$ at the ground state. As can be seen, as $\Delta Q_{20}$ increases above  $\sim$75~b, the difference between reconstructed half-lives and  the reference half-lives tends to increase.

While it may be desirable in principle to improve the emulation, the nuclei whose $\tsf$ values are reproduced with a large error are those predicted to be stable to SF, within the $(Q_{20},Q_{30})$ collective space. As such, errors in the SF observables have little effect on results that are further dependent on $\tsf$, such as $r$-process network calculations.

The inset panels in Fig.~\ref{fig:lifetime-diag} magnify the range $10^{-5}-10^{10}$~s, to highlight the relevant $r$-process range. As can be seen, almost all nuclei within this range are reproduced nicely within three orders of magnitude. Therefore, we conclude that NNs are able to reproduce both the PES and the collective inertia well enough that $\tsf$ is reproduced within 3 orders of magnitude for nuclei for which SF is relevant in the $r$-process region.

\section{Conclusions}
\label{sec:conclusion}

In this work, we have shown that fully connected feedforward  NNs are able to emulate both the potential energy and the collective inertia across a region of the nuclear chart, in the collective space consisting of the axial quadrupole and octupole moments. In general, the emulation error on the  potential energy is  about 500 keV, and the largest discrepancies are found in high-energy regions far from the fission path. The  inertia tensor is reproduced within roughly an order of magnitude. We find that the NN performance is stable with respect to changes in the architecture, while the rescaling of input variables produces a general improvement overall. Most of the exit points predicted by the NN agree with the DFT predictions within a $(\Delta Q_{20}, \Delta Q_{30})= (2\,\textrm{b}, 1\,\textrm{b}^{3/2})$ range. The SF half-lives are usually reproduced within a factor 10$^3$ over a span of more than 70 orders of magnitude. We find that the largest source of discrepancies is the emulation of the collective inertia tensor, due to the rapid changes of the  inertia tensor in regions where single-particle level crossings are present. For some very long-lived nuclei, the associated  error accumulates along the wider fission barrier. Conversely, in nuclei where fission can be a major decay mode, the emulations are in very good agreement with reference data.

Emulation error in the training dataset suggests that modifications to the training methodology may be warranted. To improve NN performance on exit points, one could introduce a hybrid loss function that includes the error in the exit point. This will likely focus the NN on the region relevant to fission. However, such a loss function requires many NEB calculations at each iteration in the training, making the NN training time unreasonable. To improve the NN performance on $\mathcal{M}_{23}$, one could train one NN to fit the (logarithm of the) magnitude of $\theta$, and another to classify the sign of $\theta$. Doing so risks misclassifying $\theta$, potentially leading to a larger error in $\mathcal{M}_{23}$. One could also experiment with different transformations of the angle, $\theta$. Given that both the half-lives and exit points are emulated reasonably well despite the misprediction of $\theta$, neither approach has been pursued in this work.

In future studies, we plan to include more collective coordinates. Dynamic pairing fluctuations have been shown to be important for accurately predicting SF half-lives~\cite{Giuliani2013,Rodriguez-Guzman2014a,Sadhukhan2014,Giuliani2014,Zhao2016}, and other multipole moments have a significant impact on both the collective motion from the ground state to the fission isomer~\cite{Sheikh2009,Staszczak2011} and the fragment yields~\cite{Warda2002,Younes2009,Zdeb2021}. The NNs described in this study are expected to have similar performance when including more collective coordinates. However, these NNs require a large amount of training data, and generating such data may be prohibitively expensive for many collective coordinates. One attractive option, as demonstrated in Ref.~\cite{Lasseri2020}, is to select nuclei for use in training in an iterative manner based on the performance of the NN, so as to reduce the required number of DFT calculations.

Another option is to consider different emulators that may require a smaller amount of training data. Reduced order modeling techniques have recently been used in nuclear physics to study both non-relativistic and relativistic DFT~\cite{Bonilla2022,Giuliani2023}, especially for use in uncertainty quantification. We intend to explore the usefulness of these techniques in emulating DFT results.

\section{Acknowledgements}
 This work was supported by the U.S. Department of Energy under Award Numbers DOE-DE-NA0004074 (NNSA, the Stewardship Science Academic Alliances program), DE-SC0013365 (Office of Science),  DE-SC0023175 (Office of Science, NUCLEI SciDAC-5 collaboration). and DE-SC0024586 (STREAMLINE collaboration); and by the Spanish Agencia Estatal de Investigaci{\'o}n (AEI) of the Ministry of Science and Innovation (MCIN) under grant agreements No.~PID2021-127890NB-I00 and No.~RYC2021-031880-I funded by MCIN/AEI/10.13039/501100011033 and the European Union-``NextGenerationEU''.


%
\end{document}